\documentclass[checkin,showpacs,aps,prl,onecolumn]{revtex4}

\usepackage{mathrsfs}

\newcommand{\bn}{\begin{enumerate}}
\newcommand{\en}{\end{enumerate}}
\newcommand{\ba}{\begin{eqnarray}}
\newcommand{\ea}{\end{eqnarray}}

\usepackage{graphicx,color}

\newcommand{\be}{\begin{equation}}
\newcommand{\ee}{\end{equation}}

\newcommand{\et}{{\it et al. }}
\newcommand{\ete}{{\it et al.}}

\def\prl{{ Phys. Rev. Lett. }}

\def\pra{{ Phys. Rev. A }}

\begin{document}

\newcommand{\clr}{\color{black}}








\title{Generating high-order optical and spin harmonics from
  ferromagnetic monolayers }


\author{G. P. Zhang$^*$}

 \affiliation{Department of Physics, Indiana State University,
   Terre Haute, IN 47809, USA }

\author{M. S. Si} \affiliation{Key Lab for Magnetism and Magnetic
  Materials of the Ministry of Education, Lanzhou University, Lanzhou
  730000, China}

\author{M. Murakami} \affiliation{Department of Physics, Indiana
  State University, Terre Haute, Indiana 47809, USA}

\author{Y. H. Bai}

\affiliation{Office of Information Technology, Indiana State
  University, Terre Haute, IN 47809, USA }

\author{Thomas F. George}

\affiliation{Office of the Chancellor
  \\Departments of Chemistry \& Biochemistry and Physics \& Astronomy
  \\University of Missouri-St. Louis, St.  Louis, MO 63121, USA }

\date{\today, Published in {\bf Nature Communications} Vol. {\bf 9}, 3031 (2018).}

\begin{abstract}
  {
High-order harmonic generation (HHG) in solids has entered a new phase
    of intensive research, with envisioned band-structure mapping on
    an ultrashort time scale. This partly benefits from a flurry of
    new HHG materials discovered, but so far has missed an important
    group. HHG in magnetic materials should have profound impact on
    future magnetic storage technology advances.  Here we introduce
    and demonstrate HHG in ferromagnetic monolayers. We find that HHG
    carries spin information and sensitively depends on the
    relativistic spin-orbit coupling; and if they are dispersed into
    the crystal momentum ${\bf k}$ space, harmonics originating from
    real transitions can be ${\bf k}$-resolved and carry the band
    structure information.  Geometrically, the HHG signal is sensitive
    to spatial orientations of monolayers.  Different from the optical
    counterpart, the spin HHG, though probably weak, only appears at
    even orders, a consequence of SU(2) symmetry. Our findings open an
    unexplored frontier -- magneto-high-order harmonic generation.
}
\end{abstract}




 \maketitle



\newcommand{\ik}{i{\bf k}}
\newcommand{\jk}{j{\bf k}}


High-order harmonic generation (HHG) in atoms and small molecules has
garnered attentions over several decades. It allows one to generate a
table-top light source with energy up to x-ray regimes and time scales
down to several hundred attoseconds
\cite{brabec,corkum2007,krausz}. This leads to the advent of
attosecond physics {\color{black}\cite{krausz2009}}, where electron
dynamics is probed on its intrinsic time scale
{\color{black}\cite{kruchinin}}.  Farkas and coworkers \cite{farkas}
were the first to generate 5th-order harmonics from a gold surface by
a picosecond laser pulse. von der Linde \et \cite{linde} reported up
to 15th order in an Al film and 14th order in glass. Theoretically
Plaja and Roso-Franco \cite{plaja} examined the mechanism of harmonic
generation in silicon, while Faisal \et \cite{faisal} developed a
nonperturbative Floquet-Bloch theory to control HHG through interband
resonances. In 2005, we predicted HHG in $\rm C_{60}$ theoretically
\cite{prl05,pra06} (see other references cited there), and Ganeev \et
\cite{ganeev2009a,ganeev2009b} experimentally demonstrated HHG in
fullerenes.  However, nanostructures \cite{ganeev2013} are
traditionally unfamiliar to researchers in atomic HHG \cite{ciappina}.
In 2011, HHG in ZnO reported by Ghimire \et \cite{ghimire2011} renewed
the interest in solid state HHG, which has quickly expanded into
monolayer \cite{yoshikawa} and multilayer graphene \cite{bowlan}, MgO
\cite{Tancogne-Dejean2017a,you2017}, Si \cite{Tancogne-Dejean2017b},
$\rm MoS_2$ \cite{liu}, $\rm Bi_2Se_3$ \cite{avetissian}, SiO$_2$
\cite{luu2015,garg2016}, Ar/Kr solids \cite{nd}, GaSe
\cite{schubert2014,langer2017,hohenleutner}, and metal-sapphire
nanostructures \cite{han}.  However, to this end, little attention has
been paid to magnetic systems.


Here we show that a single laser pulse is capable of generating
high-order harmonic generation in iron monolayers. Different from
nonmagnetic materials, the harmonic signal carries the spin
information. The majority and minority spins generate different
harmonics.  In contrast to HHG in atoms and small molecules,
circularly polarized light can generate even higher-order harmonics,
which are helicity-dependent. We compare two different Fe(110) and
Fe(001) surfaces and find that the harmonics from Fe(110) are
stronger. We find that the different density of states around the
Fermi level is responsible for this difference. We disperse harmonics
in the crystal momentum space, and we find that, in general, harmonics,
which result from virtual transitions, appear symmetric with respect
to the harmonic order and carry no information on the band
structure. However, if harmonics originate from real transitions, they
can be attributed to a few specific transitions between band states
and are useful for band structure reconstruction.  Higher harmonics
show a stronger band dispersion. Different from the charge
counterpart, due to the SU(2) symmetry, the harmonics from spin appear
at even orders. Our study opens a new direction by extending
high-harmonic generation to magnetic materials.


\vspace{0.5cm}

{\bf \large Results}

\begin{figure}
  \includegraphics[angle=0,width=1\columnwidth]{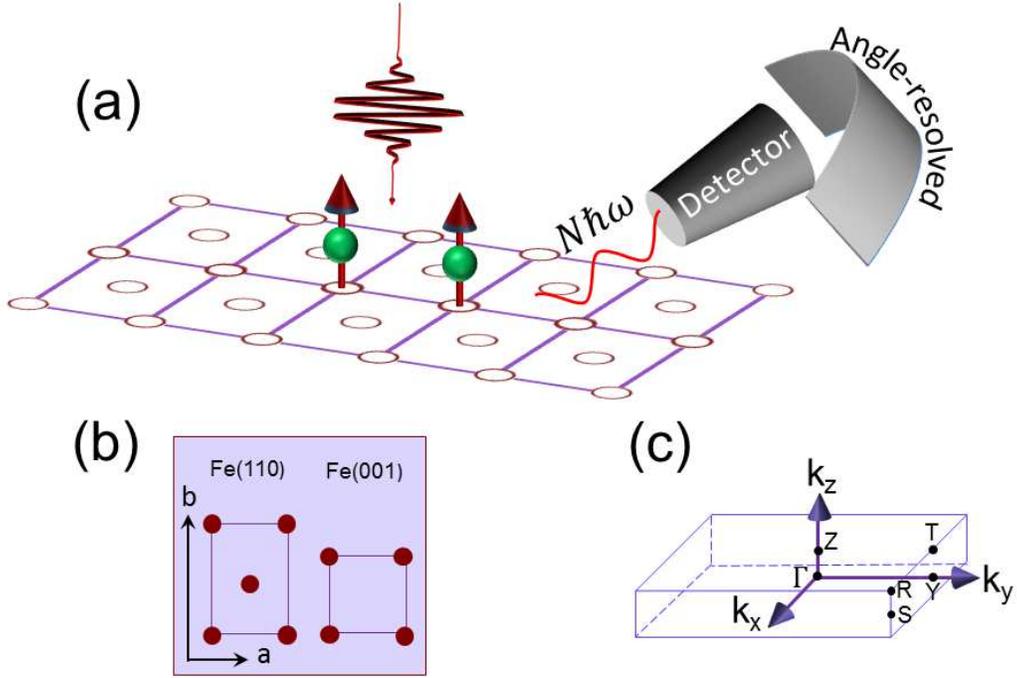}
  \caption{High-order harmonic generation in ferromagnetic materials. 
{\bf a} An intense laser pulse excites a ferromagnetic
    iron monolayer and generates high-order harmonics.  The harmonic
    has spin signature on the spectrum and can be dispersed in the
    crystal momentum space, so the harmonic peak can be assigned to a
    unique transition between occupied and unoccupied
    states. Higher-order harmonics are more sensitive to the band
    structure change than lower-order ones, thus making HHG an ideal
    tool for spin-resolved detection.  {\bf b} Brillouin zone of a
    simple orthorhombic structure.  {\bf c} Two film orientations --
    Fe(110) and Fe(001) -- are placed in the $ab$ (or $xy$) plane.  }
\label{fig1}
\end{figure}

{\color{black}{\sf \large Symmetry properties of magneto-high-order
    harmonic generation (MHHG). } HHG in nonmagnetic materials is only
  subject to the spatial symmetry.  MHHG in magnetic materials is very
  different, where spin polarization and spin-orbit coupling break the
  spatial symmetry and introduce new phenomena that are otherwise
  undetectable. This already occurs in magneto-optics, such as the
  Faraday or Kerr effect.  For instance, a cubic structure, if its
  magnetization is placed along the $z$ axis, becomes a tetragonal
  structure, and the number of symmetry operations is reduced from 48
  to 16.  Different from polar vectors, magnetic moment vectors ${\bf
    M}$ are axial vectors and transform as \be T_{\rm axial}(O){\bf
    M}={\rm det}[O] O {\bf M}, \label{spin} \ee where ${\rm det}$ is
  the determinant of the symmetry operation $O$, and $T$ is the
  transformation. There is no difference between the polar and axial
  vectors for proper rotations, but for improper rotations, they
  differ by a sign change determined by the above determinant.

For an orthorhombic system with the magnetization along the $z$ axis
(Fig. \ref{fig1}(a)), there are eight symmetry operations (see
Methods), but only four of them keep the magnetization invariant and
are retained in the symmetry group.  These symmetry operations are
essential to our understanding of MHHG. Consider a proper rotation
$O_2$ (a twofold rotation around the $z$ axis, $\rm C_{2z}$) and an
improper rotation $O_6$ (a reflection with respect to the $yz$ plane,
$\sigma_x$):
\begin{equation}
   O_2=
  \left ( {\begin{array}{rrr}
  -1 & 0  &~~0 \\
   0 &-1  &0 \\
   0 & 0 &  1\\
  \end{array} } \right ); \hspace{2cm}
   O_6=
  \left( {\begin{array}{rrr}
  -1 & ~~0  &~~0 \\
   0 &  1  &0 \\
   0 & 0 &  1\\
  \end{array} } \right).
\end{equation}
For a nonmagnetic system, both symmetry operations appear in the point
group. If the laser field is polarized along the $x$ axis, these two
symmetry operations cancel any harmonic signal along the $y$ axis.
Now consider the same laser field incident on a magnetic sample (see
Fig. \ref{fig1}).  If the system is spin-polarized along the $z$ axis,
the point group only retains $O_2$, not $O_6$ since $O_6$ changes the
direction of the spin moment. In other words, this symmetry reduction
voids the original cancellation, so a new signal appears along the $y$
axis. On the other hand, it is easy to verify that if the laser
polarization is along the $z$ axis, there is no signal along other
directions.  The entire symmetry properties can be worked out once the
symmetry group is known.  This is the theoretical foundation of MHHG.
}

\vspace{0.3cm}

{\sf \large First-principles formalism. } {\color{black} We choose
  iron monolayers as our model system.  Fig. \ref{fig1}(b) shows two
  spatial orientations in the iron monolayers -- Fe(110) and Fe(001)
  surfaces.  They are simulated by a slab geometry,} where a vacuum
spacing separates slabs so there is little interaction between them
(for details, see the Supplementary Methods). We solve the Kohn-Sham
equation (in atomic units) \cite{prb09} to find both the eigenstates
and optical transition matrices, \be
[-\nabla^2+V_{ne}+V_{ee}+V_{xc}]\psi_{\ik}(r)=E_{\ik} \psi_{\ik}
(r), \label{ks} \ee where the terms on the left-hand side represent
the kinetic energy, nuclear-electron attraction, electron-electron
Coulomb repulsion and exchange correlation {\color{black} \cite{pbe}},
respectively.  $\psi_{\ik}(r)$ is the Bloch wavefunction of band $i$
at crystal momentum ${\bf k}$, and $E_{\ik}$ is the band energy.  We
include the spin-orbit coupling using a second-variational method in
the same self-consistent iteration {\color{black} \cite{wien2k}} and
construct the spin matrices. Once our calculation reaches
self-consistency, we investigate harmonic generations by employing the
time-dependent Liouville equation, \be i\hbar \langle
\ik|\frac{\partial \rho}{\partial t}|\jk\rangle = \langle \ik |
         [H_0+H_I,\rho]|\jk\rangle, \label{liu} \ee where $\rho$ is
         the density matrix, $H_0$ is the system Hamiltonian, and
         $H_I$ is the interaction between the system and laser field:
         $H_I=\frac{e}{m_e} \hat{{\bf P}}\cdot {\bf A}(t)$, where $-e$
         is the electron charge, $m_e$ is its mass, $\hat{{\bf P}}$ is
         the momentum operator and ${\bf A}(t)$ is the laser-field
         vector potential \cite{sm}.  We choose a Gaussian pulse with
         duration $\tau$ and photon energy $\hbar\omega$.
         {\color{black} We note that the time-dependent Liouville
           density functional theory \cite{jpcm16} is advantageous
           since it rigorously respects the Pauli exclusion principle
           that two electrons can not occupy the same spin state at
           the same time. }

 After we numerically solve the density matrix $\rho$ from
 Eq. (\ref{liu}), we can compute the expectation value of the momentum
 operator \cite{prl05,pra06} by ${\bf P}(t)=\sum_{\bf k} {\rm
   Tr}[\rho_{\bf k}(t) \hat{{\bf P}}_{\bf k}]$, where the trace is
 over band indices and {\color{black} includes interband
   contributions. We only include intraband transitions indirectly
   through the interband transition. For our current laser field
   amplitude, the crystal momentum shift is very small.}  The harmonic
 spectrum is computed by Fourier transforming ${\bf P}(t)$ into the
 frequency domain through {\color{black}\cite{prl05,mitsuko2017}}, \be
 {\bf P}(\Omega)=\int_{-\infty}^{\infty} {\bf P}(t) {\rm e}^{i\Omega
   t} {\cal W}(t) dt, \ee {\color{black}where ${\cal W}(t)$ is the
   window function (see Supplementary Note 1). All the harmonic
   spectra below use $\log_{10}|{\bf P}(\Omega)|$}.  Caution must be
 taken that the time propagation during simulation must be long enough
 to resolve fine structures in HHG spectra. {\color{black} In our
   calculation, the starting time is -600 fs, and the ending time is
   typically around 600 fs and in some cases up to 1.5 ps.} The time
 step, which determines the highest harmonic order, is 1/32 the laser
 period, {\color{black} and when the field is stronger, it is 1/64 the
   laser period.}  Both the extremely long time propagation and small
 time step ensure that our spectrum is very sharp and has a
 well-defined shape.

\vspace{0.3cm}

\begin{figure}
  \includegraphics[angle=270,width=1\columnwidth]{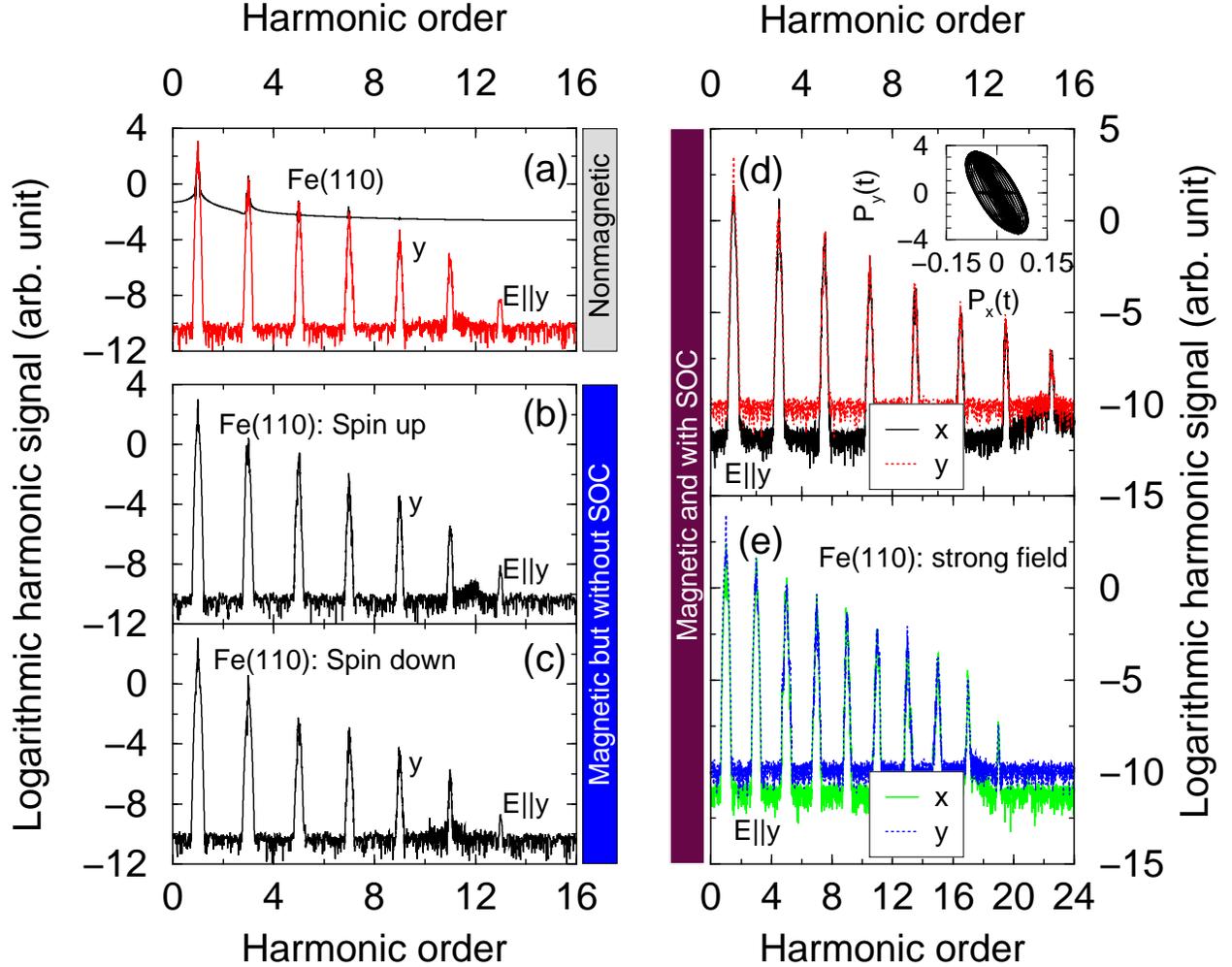}
  \caption{ Harmonic signals under different magnetic ordering and
    laser amplitudes.  {\bf a} HHG signal from a nonmagnetic Fe
    monolayer.  The laser E-field is along the $y$ axis, with
    $\hbar\omega=2.0$ eV, $\tau=60$ fs and field amplitude
    $E_0=0.09\rm V/\AA$, for the results in this figure.  The top
    curve is obtained without using the window function, while the
    bottom is processed with the window function.  {\bf b} HHG from
    the spin-up channel in a magnetic Fe monolayer. The spin-orbit
    coupling is not included.  {\bf c} Similar to ({\bf b}), but from
    the spin-down channel.  {\bf d} HHG signal with spin-polarized
    electrons and spin-orbit coupling. The solid and dashed lines
    denote the signals along the $x$ and $y$ axes, respectively.  The
    spin is orientated perpendicular to the Fe(110) surface.  Inset:
    Phase diagram of $P_x(t)$ versus $P_y(t)$.  {\bf e} Same as ({\bf
      d}) but with $E_0=0.15\rm V /\AA$, where high harmonics up to
    19th order are observed.
}
\label{fig2}
  \end{figure}

{\sf \large Spin-polarized HHG. } We employ a 60-fs linearly polarized
laser pulse along {\color{black} the $y$ axis}. Our photon energy is
$\hbar \omega=2$ eV and {\color{black} field amplitude is $0.09~\rm V
  /\AA$ (far below Bragg reflection \cite{schubert2014}).} These
parameters are commonly used in experiments {\color{black} and are
  employed for the following calculations.  We start with a
  nonmagnetic Fe(110) monolayer where we run a nonspin-polarized
  calculation. Figure \ref{fig2}(a) shows that the harmonic signals
  are only along the $y$ axis, consistent with the above symmetry
  analysis, and the highest harmonic order is 13. {\color{black} The
    top curve is the one obtained with ${\cal W}(t)=1$ and the
    bottom one with a hyper Gaussian function. } Next, we consider a
  spin-polarized case without spin-orbit coupling.  Ferromagnetic
  materials have two distinctive spin channels: majority (spin-up) and
  minority (spin-down). We carry out two separate calculations with
  the same laser parameters. Figure \ref{fig2}(b) shows the results
  for the majority spin, where for the same symmetry reason, no signal
  is found along either the $x$ or $z$ axis. The HHG signal for the
  spin-up channel also reaches up to 13th order.}  However, the
spin-down channel is quite different.  Figure \ref{fig2}(c) shows that
although its highest order is the same, the magnitude is smaller.  We
will come back to this below.  What is even more interesting is when
the spin-orbit coupling (SOC) is present. {\color{black} In this case,
  two spin channels are coupled, and the spin has a preferred spatial
  orientation, which breaks the symmetry. Figure \ref{fig2}(d) shows
  that this symmetry breaking introduces a new signal along the $x$
  axis (see the solid line), along with the ordinary harmonics along
  the $y$ axis.  This is qualitatively different from the nonmagnetic
  case where no signal is found along the $x$ axis (see
  Fig. \ref{fig2}(a)). The inset shows the real time evolution of
  $P_x(t)$ and $P_y(t)$. It is very interesting that similar to the
  time-resolved magneto-optical Kerr effect (TRMOKE) \cite{np09},
  these two components have a clear phase relation, and the major axis
  of the eclipse formed by $P_x(t)$ and $P_y(t)$ tilts slightly away
  from the $y$ axis. In TRMOKE, the angle that the major axis makes
  with the $x$ axis sensitively reflects the strength of the
  spin-orbit coupling. A similar feature observed here will be
  investigated in the future.  To be sure that the HHG signal is
  indeed from the laser field, we increase {\color{black} the field
    amplitude to 0.15 $\rm V/\AA$}, and find that the harmonic order
  increases all the way up to 13th order (see
  Fig. \ref{fig2}(e)). This demonstrates that our results are
  robust. If we compare those high harmonics with the low harmonics,
  we find that their signals do not drop significantly, so they should
  be measurable. Experimentally, the second order harmonic was already
  observed \cite{eric2004}}.  This constitutes our first major finding
that HHG in magnetic materials is spin-channel dependent and is
affected by spin-orbit coupling, a unique feature that is not shared
with other materials.

\vspace{0.3cm}

\begin{figure}
  \includegraphics[angle=270,width=1\columnwidth]{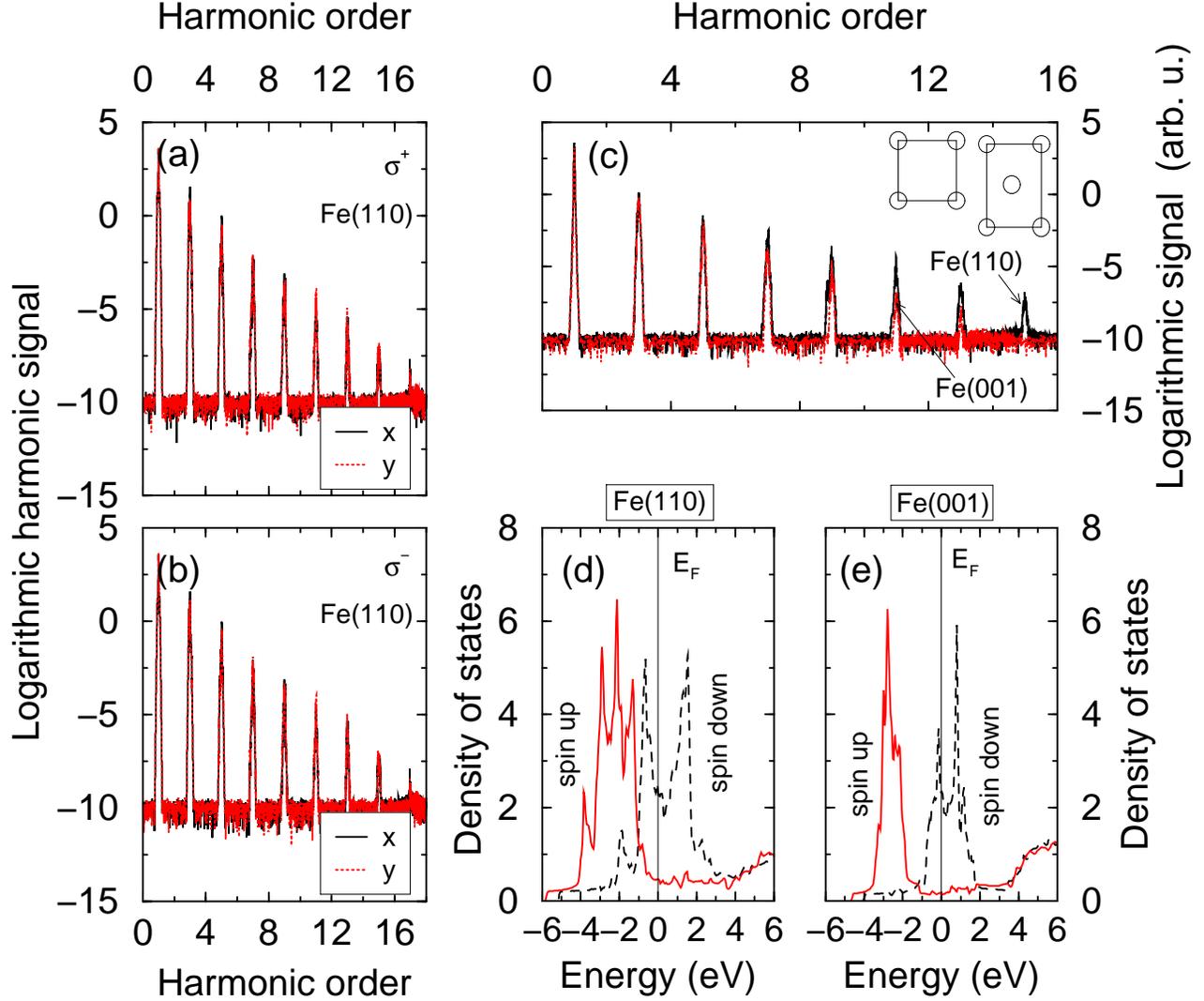}
  \caption{ Effects of laser-helicity and ï¬lm orientation on
    harmonic signals.  {\bf a} and {\bf b} Logarithmic harmonic signal
    from the Fe(110) monolayer with right ($\sigma^+$) and left
    ($\sigma^-$) circularly polarized light, respectively, where the
    laser polarization is in the $xy$ plane (see Fig. \ref{fig1}).
    Due to the window function, the difference between the $x$ and $y$
    components is not obvious, but the real time $P_y(t)$ is larger
    than $P_x(t)$ for most of the time (see Supplementary Figure 3).
    {\bf c} Comparison between HHG signals in the Fe(110) and Fe(001)
    monolayers, where the laser polarization is along the $z$ axis.
    {\bf d} Density of states for the Fe(110) monolayer.  The solid
    and dashed lines denote the spin-up and spin-down density of
    states, respectively. The Fermi level is at $E_F=0$ eV (see the
    thin vertical line).  {\bf e} Density of states for the Fe(001)
    monolayer.
  }
\label{fig3}
  \end{figure}

{\sf \large Helicity and surface orientation dependence. } Different
from HHG in atoms \cite{brabec}, we find that circularly polarized
light \cite{you2017} can effectively generate HHG in magnetic systems
as well.  We choose right ($\sigma^+$) and left ($\sigma^-$)
circularly polarized light in the $ab$ plane (see Fig. \ref{fig1}).
In traditional magneto-optics, because of the spin-orbit coupling and
exchange interaction \cite{np09}, $\sigma^+$ and $\sigma^-$ do not
generate identical signals {\color{black} because they choose
  different sets of transitions among band states.}  Figures
\ref{fig3}(a) and \ref{fig3}(b) show that HHG retains this difference,
consistent with the prior magnetic second-order harmonic generation
\cite{jorg}.  Energetically, our highest harmonic energy is
significantly higher than that in native graphene
\cite{yoshikawa,bowlan}.


There are additional knobs that one can turn in monolayers. They can
be cut along different axes, so that the crystal orientation plays a
role \cite{langer2017}.  Fe monolayers have two possible orientations,
Fe(110) and Fe(001) surfaces.  Figure \ref{fig3}(c) shows that HHG in
the Fe(110) monolayer is stronger than that in the Fe(001) monolayer;
and this remains true even with a denser $k$ mesh \cite{sm}. Such
orientational dependence has been reported before in GaSe
\cite{langer2017} and $a$-cut (11-20) ZnO \cite{gholam2017}, but never
in a magnet.  You \et \cite{you2017} explained the orientation
dependence in insulating MgO through the interatomic bonding. We do it
differently, as electrons in our system are very delocalized.

One obvious explanation is that the number of atoms in the primitive
cell is different for Fe(110) and Fe(001), but this is not the entire
story. We notice that the harmonic amplitude ratio between Fe(110) and
Fe(001) is not proportional to the atom number ratio. {\color{black}
  For instance, the ratio in the $z$ component is 2.33 for the first,
  2.93 for the third, 3.34 for the 5th, 12.14 for the 7th, 38.01
  for the 9th, and 28.05 for the 11th order. There is no signal at the
  15th harmonic for Fe(001).}
We decide to examine the density of
  states (DOS) for the majority and minority bands.  Figure
  \ref{fig3}(d) shows the total DOS in Fe(110) for the majority and
  minority states around the Fermi level $E_f$ (vertical thin line).
  The majority channel has more electrons than those in the minority
  channel, so it contributes more to harmonic generation. This
  explains the difference seen in Figs. \ref{fig2}(b) and
  \ref{fig2}(c).  We see that due to the low symmetry in Fe(110), the
  majority valence electrons have three broad peaks, and the minority
  ones also have a large DOS below $E_f$.  By contrast, the Fe(001)
  monolayer is quite different. Figure \ref{fig3}(e) shows that its
  majority band is much narrower than that in the Fe(110) monolayer,
  centered around 2.8 eV below the Fermi level.  The band is less
  dispersive than for the Fe(110) monolayer, so many channels are not
  available to the Fe(001) monolayer to generate harmonics, which
  weakens HHG in the Fe(001) monolayer (see \cite{sm} for more).




\vspace{0.5cm}

{\bf \Large Discussion}

\begin{figure}
  \includegraphics[angle=270,width=1\columnwidth]{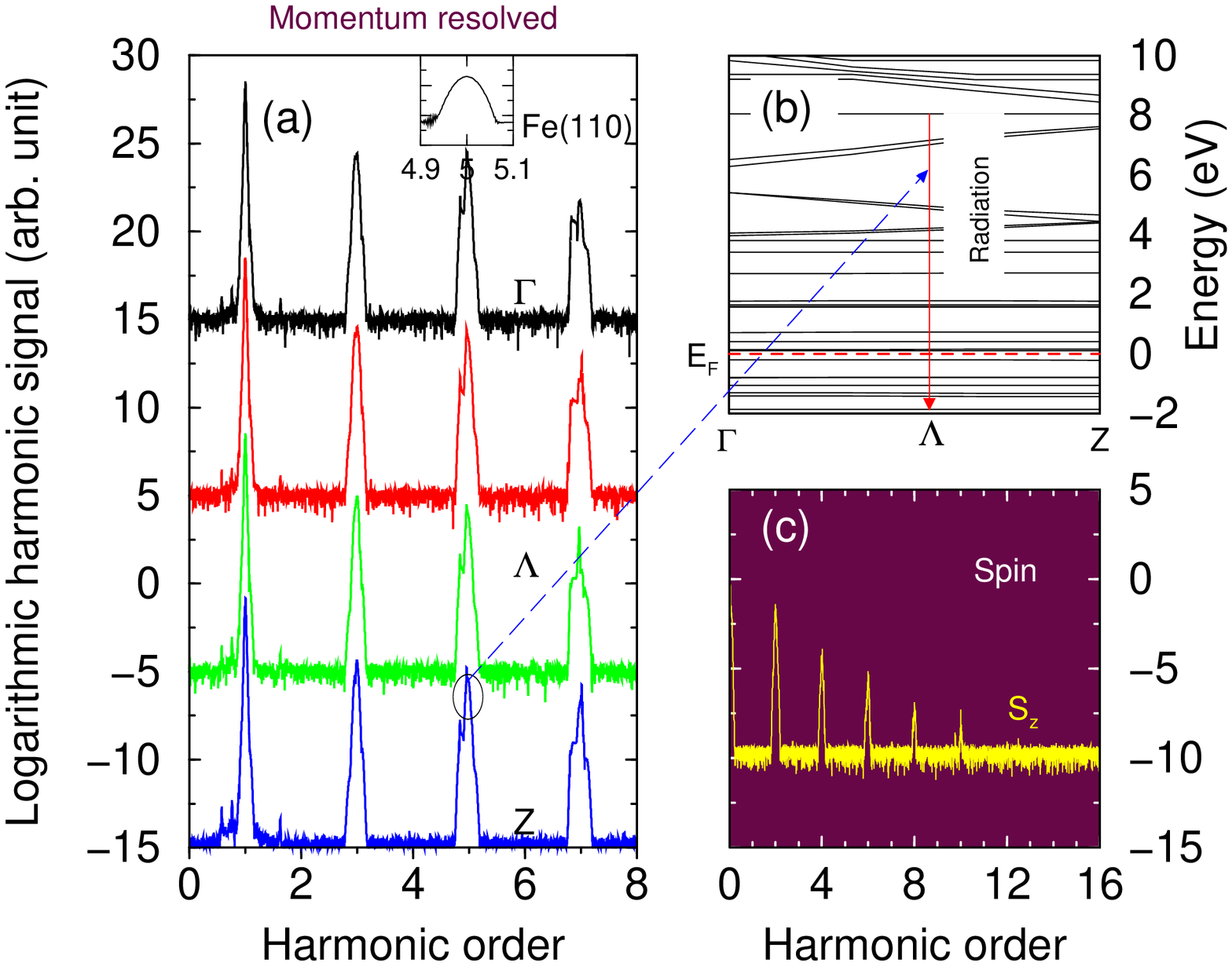}
  \caption{ Crystal-momentum-resolved high-harmonic generation and
    spin harmonic generation.  {\bf a} Crystal-momentum-resolved
    harmonic signal from the $Z$ to $\Gamma$ point on a logarithmic
    scale.  The laser is linearly polarized along the $y$ axis, and
    the signal is from the $x$ axis.  Except the one at $Z$, all the
    curves are shifted vertically.  Top inset: Zoomed-in view of the
    5th harmonic at a different $k$ point.  {\bf b} Band dispersion
    along the $\Gamma-Z$ direction for the Fe(110) monolayer.  The 5th
    harmonic corresponds to several crucial transitions from
    conduction states between 8 and 9.36 eV above the Fermi level (see
    the horizontal dashed line) to a state at -1.87 eV below the Fermi
    level. The arrow denotes this radiation. {\bf c} Spin harmonic
    generation spectrum. Its zeroth order denotes the demagnetization,
    while higher orders represent the oscillation. The harmonic only
    appears at even orders, due to SU(2). The laser field has to
    interact with the system an even number of times to affect spins.
    These signals are normally weaker than the emission from the
    electric dipole.
  }
\label{fig4}
  \end{figure}

Such a sensitive dependence of harmonic signals on DOS found here is
important, as it suggests a potentially useful application to map band
states through HHG in magnetic materials.  This reminds us of our
earlier work on $\rm C_{60}$ \cite{prl05}, where nearly every harmonic
can be uniquely assigned to a particular transition. In ZnO, Vampa \et
\cite{vampa2015a} proposed to reconstruct the band structure from the
HHG spectrum.  Experimentally, Luu \et \cite{luu2015} showed that the
EUV radiation allowed them to probe the conduction band dispersion in
SiO$_2$. A similar result was found in rare-gas solids \cite{nd}. But
none of these studies tested magnetic materials.  In the following, we
demonstrate a highly accurate yet challenging detection scheme that
the band transition states can be probed through HHG in the crystal
momentum space. We take Fe(110) as an example. We disperse the
harmonic signal in the crystal momentum space. There are many possible
pockets in the Brillouin zone that we can investigate.  Here we choose
the $\Lambda$ line that links the $\Gamma$ and $Z$ points (see
Fig. \ref{fig1}(c)).  {\color{black} The harmonic signal is dispersed
  along the $\Gamma$-Z direction, $\Lambda$ line from top to bottom in
  Fig. \ref{fig4}(a). Note that in our calculation all $\Gamma$ and
  $Z$ points are approximate since we have to shift the $k$ mesh
  slightly in order to get better convergence. For clarity, in
  Fig. \ref{fig4}(a) we vertically shift all the curves except the
  bottom one.}  These spectra carry rich information about the band
states.  We see that harmonics at different orders change with $k$
differently and this change is not limited to the lower-order
harmonics.  Higher-order harmonics show an even stronger dispersion.
We take the fifth harmonic as an example. We see there are many
smaller peaks. These peaks do not distribute symmetrically around the
nominal 5th order.  {\color{black} This is an important indication
  that the harmonic engages real transitions among band states
  \cite{prl05}, where the harmonic carries the band structure
  information and thus allows one to crystal-momentum resolve the
  bands. }

Through a tedious but straightforward procedure (see details in the
Supplementary Note 2), we pinpoint the origin of the 5th harmonic.  It
comes from radiation (see the arrow in Fig. \ref{fig4}(b)) from five
conduction-band states between 8.0 and 9.36 eV above $E_f$ to a
valence state around -1.87 eV below the Fermi level $E_f$ (dashed
line).  {\color{black}Therefore, interband transitions dominate the
  spectrum. } We verify that if we remove these transition states, the
5th harmonic reduces sharply.  The largest transition matrix element
for the 5th harmonic is $(-0.4-i 0.068) 10^{-2}\hbar\rm/bohr$.
{\color{black} To develop a generic picture of the limits of the
  crystal-momentum-resolved HHG, we generate a different set of $k$
  points and compute their HHG spectra. We show one example in the top
  inset in Fig. \ref{fig4}(a), where we see a nice symmetric
  Gaussian-like distribution around the 5th order. We find that these
  symmetric peaks normally result from virtual excitations, carry no
  information about the band states, and can not be resolved in the
  crystal momentum space. Even if we systematically exclude relevant
  transitions, we can not cleanly remove the peak until we delete all
  the transitions or tune down the laser field.}  This suggests that
HHG is potentially a powerful tool to detect band transitions in the
crystal momentum space.  Such a detection scheme is achieved by three
crucial elements in HHG: (i) the incident photon energy pre-selects
dipole-allowed band states, (ii) when the HHG is dispersed into the
$k$ space, it further limits them to fewer band states, and (iii) HHG
must result from real transitions among band states.  Experimentally
one must first examine the peak structure. This ensures high accuracy
of our proposed scheme.



After we solve the Liouville
equation (\ref{liu}), the optical HHG is not the only one that we can
investigate.  We can also compute the spin change through ${\bf
  S}(t)=\sum_{\bf k} {\rm Tr}[\rho_{\bf k}(t) \hat{\bf S}_{\bf k}]$,
where $\hat{\bf S}_k$ is the spin operator, and then we
Fourier-transform it into the frequency domain. Figure \ref{fig4}(c)
shows our results. Its zeroth order is the baseline of spin moment,
and reflects how much the laser pulse demagnetizes our spin
system. Demagnetization is only part of the entire process, and spin
also oscillates with time. Our spectrum surely catches this, but the
harmonic peak only appears at even orders. This is because the spin
has SU(2) symmetry, and the laser field must interact with the system
at least twice to affect the spin. $S_z$ is dominated by the zeroth
order.  {\color{black} We do not find a higher-order harmonic beyond
  the 10th order with our current laser parameters. We caution that
  emission from spin in general is much weaker. However, with new
  experimental developments \cite{schubert2014,hohenleutner}, these
  signals should be detectable. One advantage in systems with
  inversion symmetry is that the dipole radiation has no even
  harmonic, so the emission from spin is essentially background free.
}  Nonlinear magneto-optical investigations in ferromagnetic
monolayers and thin films have a long history.  Second-harmonic
generation has been extensively used to probe surface magnetism
\cite{regensburger2000,jorg}.  High-harmonic generation has already
been used to probe ultrafast and element-specific magnetization
\cite{chan2009,mathias2012}, and THz emission from magnetic thin films
was reported \cite{eric2004,huisman2017}.  Thus, our findings are
likely to motivate further investigations in the future \cite{sm}.

{\bf \Large Acknowledgments } This work was supported by the
U.S. Department of Energy under Contract No. DE-FG02-06ER46304 (GPZ,
MM, and YHB). MSS acknowledges the support of the NSFC of China Under
Nos. 51372107 and 11774139.  Part of the work was done on Indiana
State University's high performance quantum and obsidian clusters.
The research used resources of the National Energy Research Scientific
Computing Center, which is supported by the Office of Science of the
U.S. Department of Energy under Contract No. DE-AC02-05CH11231.



{\color{black}
{\bf \Large Methods}

{\sf \large Time-dependent Liouville density functional theory. ~~} Our
theoretical calculation consists of two steps. First, we solve the
Kohn-Sham equation \cite{wien2k} to obtain the eigenvalues and
eigenstates.  We employ the generalized gradient approximation at the
PBE level \cite{pbe} as implemented in the Wien2k code. The code
employs the full-potential linearlized augmented plane-wave method,
where dual basis functions are used in the atomic sphere and
interstitial regions, and no approximation to the sphere is made. This
makes calculations very accurate. The product of the Muffin-tin radius
and plane-wave cutoff is $R_{MT}K_{max}=7$. In the spin-orbit coupling
calculation, we use a large orbital angular momentum quantum number of
$L=6$ to ensure the high accuracy of the spin matrices; and all the
eigenstates up to 3.5 Ryd are computed.  This is the same maximum
energy used in the optical calculation where transition matrix
elements are computed. We have changed the original optic code so we
can store all those matrices in an unformatted form, which improves
the accuracy of the HHG calculation greatly.

Next we solve the Liouville equation for density matrices in the time
domain for all the $k$ points. This step is most time-consuming since
we have to solve thousands of equations simultaneously. Our code is
fully parallelized using the MPI architecture.  Once we find the
density matrices at each time step, we compute the expectation value
of the momentum operator or other interesting quantities by tracing
all the product of density matrices and an operator $\mathscr{O}$,
i. e.  $\sum_k{\rm Tr}(\rho_k(t) \mathscr{O})$. Here $\rho$ depends on
the space group symmetry through $H_I$, for which we show one example
below.

{\sf \large Symmetry analysis in an orthorhombic lattice. ~} The symmetry
group, for a nonmagnetic orthorhombic lattice as well as for a
magnetic orthorhombic lattice without spin-orbit coupling, includes
all eight symmetry operations, $\{O_i\}$ where $i$ runs from 1 to 8:
\begin{equation}
O_1=\left(
\begin{array}{rrr}
-1 & 0 & 0\\
 0 &-1 & 0\\
 0 & 0 & -1\\
\end{array}
\right ),
O_2=
\left(
\begin{array}{rrr}
-1 & 0 &~~ 0\\
 0 &-1 & 0\\
 0 & 0 & 1\\
\end{array}
\right ),
O_3=\left(
\begin{array}{rrr}
1 & ~~0 & 0\\
0 &1 & 0\\
 0 & 0 & -1\\
\end{array}
\right ),
O_4=\left(
\begin{array}{rrr}
1 & ~~0 & 0\\
 0 &1 & ~~0\\
 0 & 0 & 1\\
\end{array}
\right )
\label{sym1}
\end{equation}

\begin{equation}
O_5=\left(
\begin{array}{rrr}
-1 & ~~0 & 0\\
 0 &1 & 0\\
 0 & 0 & -1\\
\end{array}
\right ),
O_6=
\left(
\begin{array}{rrr}
-1 & ~~0 & ~~0\\
 0 &1 & 0\\
 0 & 0 & 1\\
\end{array}
\right ),
O_7=\left(
\begin{array}{rrr}
1 & 0 & 0\\
0 &-1 & 0\\
 0 & 0 & -1\\
\end{array}
\right ),
O_8=
\left(
\begin{array}{rrr}
1 & 0 & ~~0\\
 0 &-1 & 0\\
 0 & 0 & 1\\
\end{array}
\right ) .
\label{sym2}
\end{equation}

\noindent In general, there are eight different density matrices,
$\rho_k\left [(e/m)O_i\hat{\bf P}\cdot {\bf A}(t)\right ]\equiv
\rho_k(O_i)$. The symmetry operation also applies to the operator
$\mathscr{O}$ as $O_i\mathscr{O}$, so the symmetrized trace is
$\sum_{i=1,8}\sum_k {\rm Tr}\left [ \rho_k(O_i) (O_i\mathscr{O})\right
]$. However, for a spin-polarized and spin-orbit coupled system, only
the first four operations ($O_1$, ..., $O_4$) remain in the group
(Eq. (\ref{sym1})), while Eq. (\ref{sym2}) is left out because they
change the spin direction (Eq. (\ref{spin})). This is the origin of
the magneto-high-harmonic generation. For other systems, one can use
the same method to work out the details.  }

{\sf\large Data availability.}  The data that support the plots within
this paper and other findings of this study are available from the
corresponding author upon reasonable request.

\vspace{1cm}

{\bf \Large Author contributions. } GPZ and MSS designed and carried
out the calculation. GPZ drafted the paper with contributions from all
the authors. All the authors discussed the results. YHB parallelized
the code and ported it to NERSC.

\vspace{0.5cm}

{\bf \Large Competing interests.}  The authors declare
  no competing interests.

\vspace{0.5cm}

{\bf \Large Materials and Correspondence. } All correspondence and
material requests should be addressed to GPZ at
$^*$gpzhang.physics@gmail.com.

\begin{thebibliography}{99}

\bibitem{brabec}T. Brabec and F. Krausz, {Intense few-cycle laser
fields: Frontiers of nonlinear optics}, Rev. Mod. Phys.  {\bf 72},
545 (2000).

\bibitem{corkum2007}P. B. Corkum and F. Krausz, {Attosecond science},
Nat. Phys. {\bf 3}, 381 (2007).

\bibitem{krausz}F. Krausz and M. I. Stockman, {Attosecond metrology
from electron capture to future signal processing},
Nat. Photon. {\bf 8}, 205 (2014).

\bibitem{krausz2009} F. Krausz and M. Ivanov, {Attosecond physics},
Rev. Mod. Phys. {\bf 81}, 163 (2009).

\bibitem{kruchinin}S. Y. Kruchinin, F. Krausz and V. S. Yakovlev,
{Colloquium: Strong-field phenomena in periodic systems},
Rev. Mod. Phys. {\bf 90}, 021002 (2018).

\bibitem{farkas} Gy. Farkas, Cs. T\'{o}th, S. D. Moustaizis,
N. A. Papadogiannis, and C. Fotakis, {Observation of
multiple-harmonic radiation induced from a gold surface by
picosecond neodymium-doped yttrium aluminum garnet laser pulses},
Phys. Rev. A {\bf 46}, R3605 (1992).

\bibitem{linde} D. von der Linde, T. Engers, G. Jenke, P. Agostini,
G. Grillon, E. Nibbering, A. Mysyrowicz and A. Antonetti,
{Generation of high-order harmonics from solid surfaces by intense
femtosecond laser pulses}, Phys. Rev. A {\bf 52}, R25 (1995).

\bibitem{plaja}L. Plaja and L. Roso-Franco, {High-order harmonic
generation in a crystalline solid}, Phys. Rev. B {\bf 45}, 8334
(1992).

\bibitem{faisal} F. H. M. Faisal and J. Z. Kami\'{n}ski, {Generation
and control of high harmonics by laser interaction with transmission
electrons in a thin crystal}, Phys. Rev. A {\bf 54}, R1769 (1996).

\bibitem{prl05} G. P. Zhang, {Optical high harmonic generations in
$\rm C_{60}$}, \prl {\bf 95}, 047401 (2005).

\bibitem{pra06} G. P. Zhang and T. F. George,
{Ellipticity dependence of optical harmonic generation in $\rm
C_{60}$}, \pra {\bf 74}, 023811 (2006).

\bibitem{ganeev2009a} R. Ganeev, L. Bom, J. Abdul-Hadi, M. Wong,
J. Brichta, V. Bhardwaj, and T. Ozaki, {Higher-order harmonic
generation from fullerene by means of the plasma harmonic method},
Phys. Rev. Lett. {\bf 102}, 013903 (2009).

\bibitem{ganeev2009b}R. Ganeev, L. E.  Bom, M. C. H. Wong,
J.-P. Brichta, V. Bhardwaj, P. Redkin, and T. Ozaki, {High-order
harmonic generation from $\rm C_{60}$-rich plasma}, Phys. Rev. A
{\bf 80}, 043808 (2009).

\bibitem{ganeev2013} R. A. Ganeev, {High-Order Harmonic Generation in
Laser Plasma Plumes}, (Imperial College Press, London, 2013).

\bibitem{ciappina}M. F. Ciappina, J. A. Perez-Hernandez, A. S.
Landsman, W. A. Okell, S. Zherebtsov, B. F\"org, J. Sch\"otz, L.
Seiffert, T. Fennel, T. Shaaran, T. Zimmermann, A. Chacón, R.
Guichard, A. Zair, J. W. G. Tisch, J. P. Marangos, T. Witting, A.
Braun, S. A. Maier, L. Roso, M. Kr\"uger, P. Hommelhoff,
M. F. Kling, F. Krausz, and M. Lewenstein, {Attosecond physics at
the nanoscale}, Rep. Prog. Phys. {\bf 80}, 054401 (2017).

\bibitem{ghimire2011} S. Ghimire, E. Sistrunk, P. Agostini,
L. F. DiMauro and D. A. Reis, {Observation of high-order harmonic
generation in a bulk crystal}, Nat. Phys. {\bf 7}, 138
(2011).

\bibitem{yoshikawa}N. Yoshikawa, T. Tamaya, and K. Tanaka,
{High-harmonic generation in graphene enhanced by elliptically
polarized light excitation},  Science {\bf 356}, 736 (2017).

\bibitem{bowlan} P. Bowlan, E. Martinez-Moreno, K. Reimann,
T. Elsaesser, and M. Woerner, {Ultrafast terahertz response of
multilayer graphene in the nonperturbative regime}, Phys. Rev. B
{\bf 89}, 041408(R) (2014).

\bibitem{Tancogne-Dejean2017a} N. Tancogne-Dejean, O.  D. M\"ucke,
F. X. K\"artner, and A. Rubio, {Ellipticity dependence of
high-harmonic generation in solids originating from coupled
intraband and interband dynamics}, Nature Comm. {\bf 8}, 745
(2017).

\bibitem{you2017} Y. S. You, D. A. Reis and S. Ghimire, {Anisotropic
high-harmonic generation in bulk crystals}, Nat. Phys. {\bf 13}, 345
(2017).

\bibitem{Tancogne-Dejean2017b} N. Tancogne-Dejean, O. D. M\"ucke,
F. X. K\"artner, and A. Rubio, {Impact of the electronic band
structure in high-harmonic generation spectra of solids},
Phys. Rev. Lett. {\bf 118}, 087403 (2017).

\bibitem{liu} H. Liu, Y. Li, Y. S. You, S.  Ghimire, T. F. Heinz, and
D. A. Reis, {High-harmonic generation from an atomically thin
semiconductor}, Nat. Phys. {\bf 13}, 262 (2017).

\bibitem{avetissian}H. K. Avetissian, A. K. Avetissian, B. R. Avchyan
and G. F. Mkrtchian, {Multiphoton excitation and high-harmonic
generation in toplogical insulator}, Preprint at http:/arXiv.gov/1711.07840v1 (2017).

\bibitem{luu2015}T. T. Luu, M. Garg, S. Yu. Kruchinin, A. Moulet,
M. Th. Hassan, and E. Goulielmakis, {Extreme ultraviolet
high-harmonic spectroscopy of solids}, Nature {\bf 521}, 498 (2015).

\bibitem{garg2016}M. Garg, M. Zhan, T. T. Luu, H. Lakhotia,
T. Klostermann, A. Guggenmos, and E. Goulielmakis, {Multi-petahertz
electronic metrology}, Nature {\bf 538}, 359 (2016).

\bibitem{nd} G. Ndabashimiye, S. Ghimire, M. Wu, D.  A. Browne,
K. J. Schafer, M.  B. Gaarde, and D. A. Reis, {Solid-state harmonics
beyond the atomic limit}, Nature {\bf 534}, 520 (2016).

\bibitem{schubert2014} O. Schubert, M. Hohenleutner, F. Langer,
B. Urbanek, C. Lange, U. Huttner, D. Golde, T. Meier, M. Kira,
S. W. Koch and R. Huber, {Sub-cycle control of terahertz
high-harmonic generation by dynamical Bloch oscillations},
Nat. Photon. {\bf 8}, 119 (2014).

\bibitem{langer2017} F. Langer, M. Hohenleutner, U. Huttner, S. W. Koch,
M. Kira, and R. Huber, {Symmetry-controlled temporal structure of
high-harmonic carrier fields from a bulk crystal}, Nat. Photonics
{\bf 11}, 227 (2017).

\bibitem{hohenleutner}M. Hohenleutner, F. Langer, O. Schubert,
M. Knorr, U. Huttner, S. W. Koch, M. Kira, and R. Huber, {Real-time
observation of interfering crystal electrons in high-harmonic
generation}, Nature {\bf 523}, 572 (2015).

\bibitem{han}S. Han, H. Kim, Y. W. Kim, Y.-J. Kim, S. Kim, I.-Y. Park,
and S.-W. Kim, {High-harmonic generation by field enhanced
femtosecond pulses in metal-sapphire nanostructure},
Nat. Comm. {\bf 7}, 13105 (2016).

\bibitem{cox}J. D. Cox, A. Marini, and G. J. G. de Abajo,
{Plasmon-assisted high-harmonic generation in graphene},
Nat. Comm. {\bf 8}, 14380 (2017).

\bibitem{luu2016}T. T. Luu and J. J. W\"orner, {High-order harmonic
generation in solids: A unifying approach}, Phys. Rev. B {\bf 94},
115164 (2016).

\bibitem{vampa2015a}G. Vampa, T. J. Hammond, N. Thire, B. E. Schmidt,
F. Legare, C. R. McDonald, T. Brabec, D. D. Klug, and P. B. Corkum,
{All-optical reconstruction of crystal band structure},
Phys. Rev. Lett.  {\bf 115}, 193603 (2015).

\bibitem{prb09} G. P. Zhang, Y. H. Bai, and T. F. George, {Energy- and
crystal momentum-resolved study of laser-induced femtosecond
magnetism}, Phys. Rev. B {\bf 80}, 214415 (2009).

\bibitem{pbe}J. P. Perdew, K. Burke, and M. Ernzerhof, Generalized
gradient approximation made simple, Phys. Rev. Lett. {\bf 77}, 3865
(1996).

\bibitem{wien2k} P. Blaha, K. Schwarz, G. K. H. Madsen, D. Kvasnicka,
and J. Luitz, WIEN2k, An Augmented Plane Wave + Local Orbitals
Program for Calculating Crystal Properties (Karlheinz Schwarz,
Techn. Universit\"at Wien, Austria, 2001).

\bibitem{sm}See Supplementary Materials.

\bibitem{jpcm16}G. P. Zhang, Y. H. Bai, and T. F. George, {Ultrafast
reduction of exchange splitting in ferromagnetic nickel}, J. Phys.:
Condens. Mat. {\bf 28}, 236004 (2016).

\bibitem{mitsuko2017} M. Murakami, G. P. Zhang, and S.-I Chu,
{Multielectron effects in the photoelectron momentum distribution of
noble-gas atoms driven by visible-to-infrared-frequency laser
pulses: A time-dependent density-functional-theory approach},
Phys. Rev. A {\bf 95}, 053419 (2017).

\bibitem{np09}G. P. Zhang, W. H\"ubner, G. Lefkidis, Y. Bai, and
T. F. George, {Paradigm of the time-resolved magneto-optical Kerr
effect for femtosecond magnetism}, {Nat. Phys.} {\bf 5}, 499
(2009).

\bibitem{eric2004} E. Beaurepaire, G. M. Turner, S. M. Harrel,
M. C. Beard, J.-Y. Bigot, and C. A. Schmuttenmaer, Coherent
terahertz emission from ferromagnetic films excited by femtosecond
laser pulses, Appl. Phys. Lett. {\bf 84}, 3465 (2004).

\bibitem{jorg}J. P. Dewitz and W. H\"ubner, {Nonlinear magneto-optics
of freestanding Fe monolayers from first principles}, Appl. Phys. B
{\bf 68}, 491 (1999).

\bibitem{gholam2017}S. Gholam-Mirzaei, J. Beetar and M. Chini, {High
harmonic generation in ZnO with a high-power mid-IR OPA},
Appl. Phys. Lett. {\bf 110}, 061101 (2017).

\bibitem{regensburger2000} H. Regensburger, R. Vollmer, and
J. Kirschner, Time-resolved magnetization-induced second-harmonic
generation from the Ni(110) surface, Phys. Rev. B {\bf 61}, 14716
(2000).

\bibitem{chan2009}C. La-O-Vorakiat, M. Siemens, M.
M. Murnane, H. C. Kapteyn, S. Mathias, M. Aeschlimann,
P. Grychtol, R. Adam, C. M. Schneider, J. M. Shaw, H.
Nembach, and T. J. Silva, { Ultrafast demagnetization dynamics at
the M edges of magnetic elements observed using a tabletop
high-harmonic soft x-ray source}, Phys. Rev. Lett. {\bf 103}, 257402
(2009).

\bibitem{mathias2012} S. Mathias \ete, Probing the timescale of the
exchange interaction in a ferromagnetic alloy, PNAS {\bf 109}, 4792
(2012).

\bibitem{huisman2017} T. J. Huisman and Th. Rasing, {THz emission
spectroscopy for THz spintronics}, J. Phys. Soc. Jpn. {\bf 86}, 011009 (2017)

\end{thebibliography}
\end{document}